\documentstyle[11pt,IAU207_pasp,twoside,psfig]{article}
\def\edcomment#1{\iffalse\marginpar{\raggedright\sl#1\/}\else\relax\fi}
\reversemarginpar

\begin{document}

\title{Evolution of Globular Clusters Formed in Mergers}
\author{Fran\c cois Schweizer}
\affil{The Observatories of the Carnegie Institution of Washington,\\ 
813 Santa Barbara Street, Pasadena, CA 91101-1292}

\begin{abstract}
Globular clusters formed in galactic mergers (e.g., The Antennae) can now
be studied at different stages of their evolution.  In young merger remnants
(e.g., NGC 7252) these ``second-generation'' globulars appear by the
hundreds as young halo clusters of roughly solar metallicity.  While at first
bluer and much more luminous than old metal-poor globulars, they become
redder after 1\,--\,1.5~Gyr and can then be observed as still overluminous
red clusters of intermediate age in perturbed-looking E and S0 galaxies
(e.g., NGC 1316, 1700, 3610).  There is evidence from the color distributions,
projected radial distributions, and perhaps also luminosity functions that
these clusters eventually assume the properties of red metal-rich globulars
observed in many giant ellipticals.  Studies of globular clusters in ongoing
mergers and young remnants suggest that second-generation globulars form
from giant molecular clouds shocked by the rapid pressure increase in the
merger-induced starburst.  This pressure-induced formation lends credence to
Cen's (2001) argument that the general pressure increase during cosmological
reionization at $z$~$\approx$ 7\,--\,15 triggered the near-simultaneous
formation of the universal population of first-generation metal-poor globulars
observed in galaxies of all types.
\end{abstract}

\section{Introduction}

Major mergers of gas-rich spirals trigger bursts of intense star and
cluster formation. They also form remnants that in many respects resemble
elliptical galaxies. The present review concentrates on recent
progress in our understanding of the evolution of globular clusters (GCs)
formed during mergers and on the growing evidence that seems to link such
globulars to the metal-rich, second-generation GCs observed in elliptical
and S0 galaxies with bimodal cluster distributions.

The merging process is now well understood.  When two
disk galaxies with massive dark halos collide, they experience dynamical
friction that leads to orbital decay and merging (Toomre \& Toomre 1972).
The accompanying violent relaxation due to the fluctuating gravitational
field redistributes the matter into a characteristic form relatively
well approximated by a $r^{1/4}$ law.  Modern simulations of mergers of 
gas-rich disks (e.g., Barnes \& Hernquist 1996; Barnes 1998, esp.\ Plate 4)
describe in considerable detail the fate of the stars, cool gas, and gas
heated to X-ray temperatures.  The resulting model remnants share many
properties with observed ellipticals, something that no other
elliptical-formation scenario has yet achieved.  Especially striking is
the speed with which these mergers occur:  From the first close approach
of two disks to their coalescence takes only about 1\onequarter\ -- 1\onehalf\
typical disk revolutions or $\sim${1/3}~Gyr.

This speed poses a challenge when we study GCs formed during ancient
mergers.  Whereas the merger and cluster-formation processes are
essentially linear in time, our ability to determine cluster ages is
more nearly time logarithmic.  Consider that when we look back in time,
the eleven ongoing mergers and young remnants that make up Toomre's (1977)
well-known sequence cover one full decade in age, ranging from
$\sim$0.1~Gyr to 1.0~Gyr.  Therefore, we can study the cluster-formation
process along this sequence in considerable detail.  Progress is also
being made in studies of GC systems in {\it intermediate-age}\, merger
remnants, which I here define as 1\,--\,7~Gyr old remnants and 
which---therefore---cover another 85\% of an age decade [$\log\,$Age(yr)~=
9.0\,--\,9.85].  But when we study cluster systems in {\it old}\, merger
remnants that formed 7\,--\,14~Gyr ago, all details appear compressed into
a mere 30\% of one age decade, $\log\,$Age(yr)~$\approx$ 9.85\,--\,10.15,
even though this interval covers fully one half of the age of the Universe.
This explains both our difficulties in understanding the details of early
cluster formation and the resulting profusion of theories.

Logically, this review should begin with the formation and evolution of
GCs in ongoing mergers, of which The Antennae are the prototype.  Yet,
this subject is already well covered by Whitmore's and Mengel's reviews
elsewhere in this volume.  Hence, I concentrate here on describing the
evolution of GC systems in young, intermediate-age, and old merger remnants.

\section{Globular Clusters in Young Merger Remnants}

Two advantages of studying GCs in young (0.3\,--\,1~Gyr) merger remnants
are that (1) dust obscuration is much less of a problem than in ongoing
mergers and (2) most bright point-like sources are true globular clusters. 
The second fact follows from the clusters' measured half-light radii and
ages.  These ages typically exceed 100~Myr, or $\sim$25\,--\,50
internal cluster-crossing times $t_{\rm cr}$,
and thus indicate that such clusters are gravitationally bound.  In contrast,
most clusters in ongoing mergers like NGC 4038/39 and NGC 3256 are
younger than $\sim$30~Myr or $\sim$10$\,t_{\rm cr}$ and may eventually
disperse.  Therefore, the time lapse between the peak of cluster formation
and the completion of a merger helps separate the wheat (i.e., the GCs) from
the chaff.

\begin{figure}[t]
\centerline{\psfig{figure=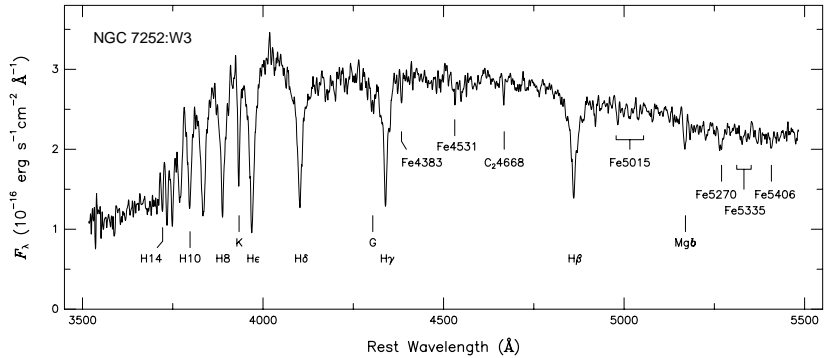,height=4.8cm}}
\caption{
UV-to-visual spectrum of young halo cluster W3 in NGC 7252.
Note strong Balmer absorption lines, the K line of Ca$\;$II, and
relatively strong metal lines around H$\beta$. This cluster has an age
of 540 $\pm 30$~Myr and metallicity $[Z] = 0.00\pm 0.08$ (Schweizer \&
Seitzer 1998).
}
\label{fig1}
\end{figure}

Globular cluster systems have been studied with the {\it Hubble Space
Telescope (HST)\,} in four young merger remnants so far:  NGC 1275, 3597,
3921, and 7252 (for references, see Table~\ref{table1}).  In each of these
remnants, there are about \mbox{10$^2$\,--\,10$^3$} point-like sources that
appear to be luminous young GCs. Age dating based on broad-band photometry
shows that the majority of these globulars formed in a relatively short,
100\,--\,200~Myr time span {\it during}\, the merger.  The young
GCs appear strongly concentrated toward their host galaxies' centers, half
of them typically lying within a central projected radius of $\sim$5~kpc.
In NGC 3921 there are also, in addition to the 102 GCs, about
50 fuzzier objects which Schweizer et al.\ (1996) called ``associations.''
These associations have colors ranging from relatively blue to quite red
and may be in the process of dispersing.  They may represent an earlier
stage of the ``faint fuzzies'' seen by Brodie (this volume) and collaborators
in three early-type galaxies.  In NGC~3921 only three of the
50 associations lie within the central 5~kpc, presumably because most
associations were too fragile to survive the intense churning at the
merger's center.

\begin{figure}[t]
\centerline{\hbox{
\psfig{figure=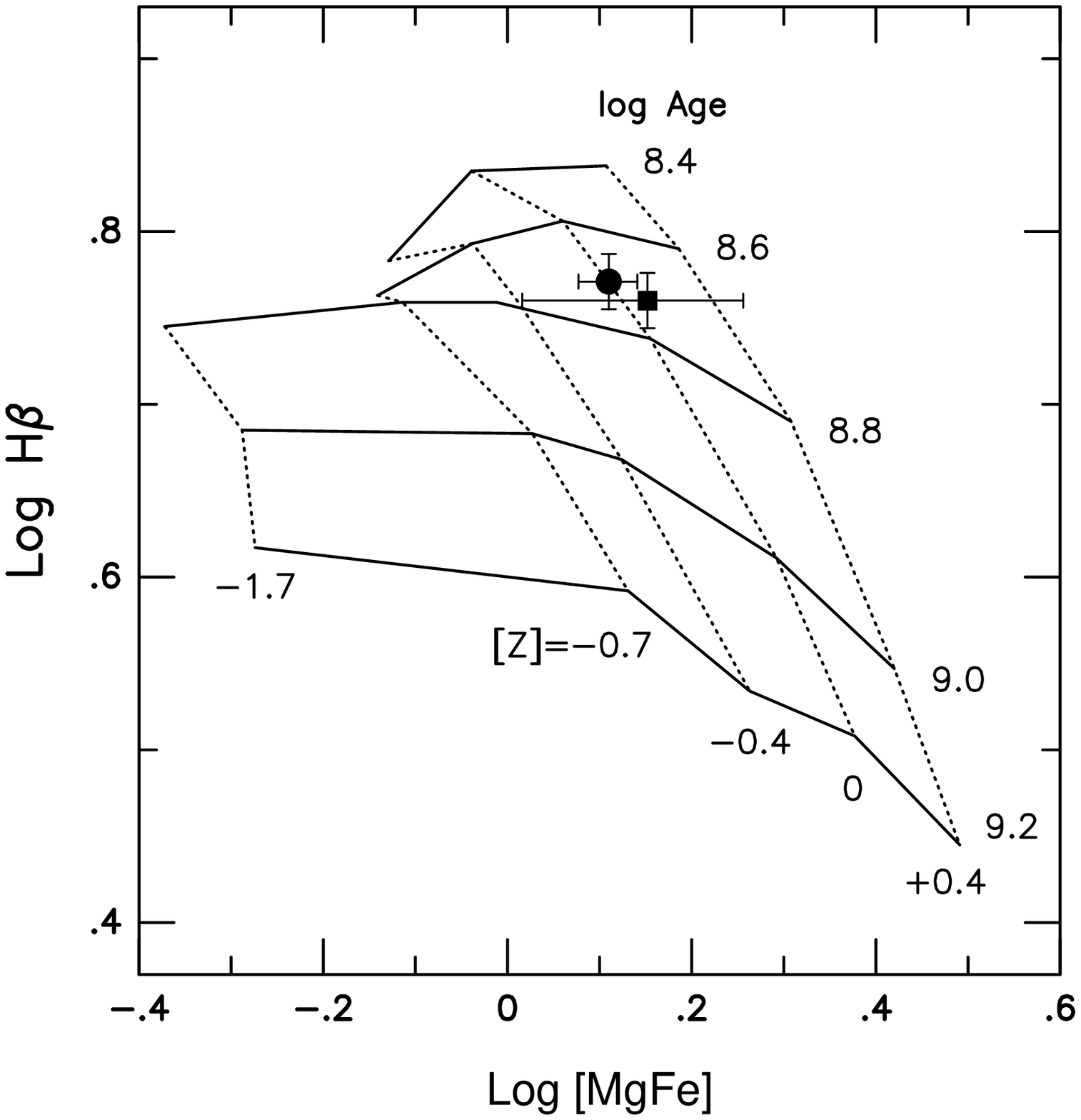,width=5.6cm}
\psfig{figure=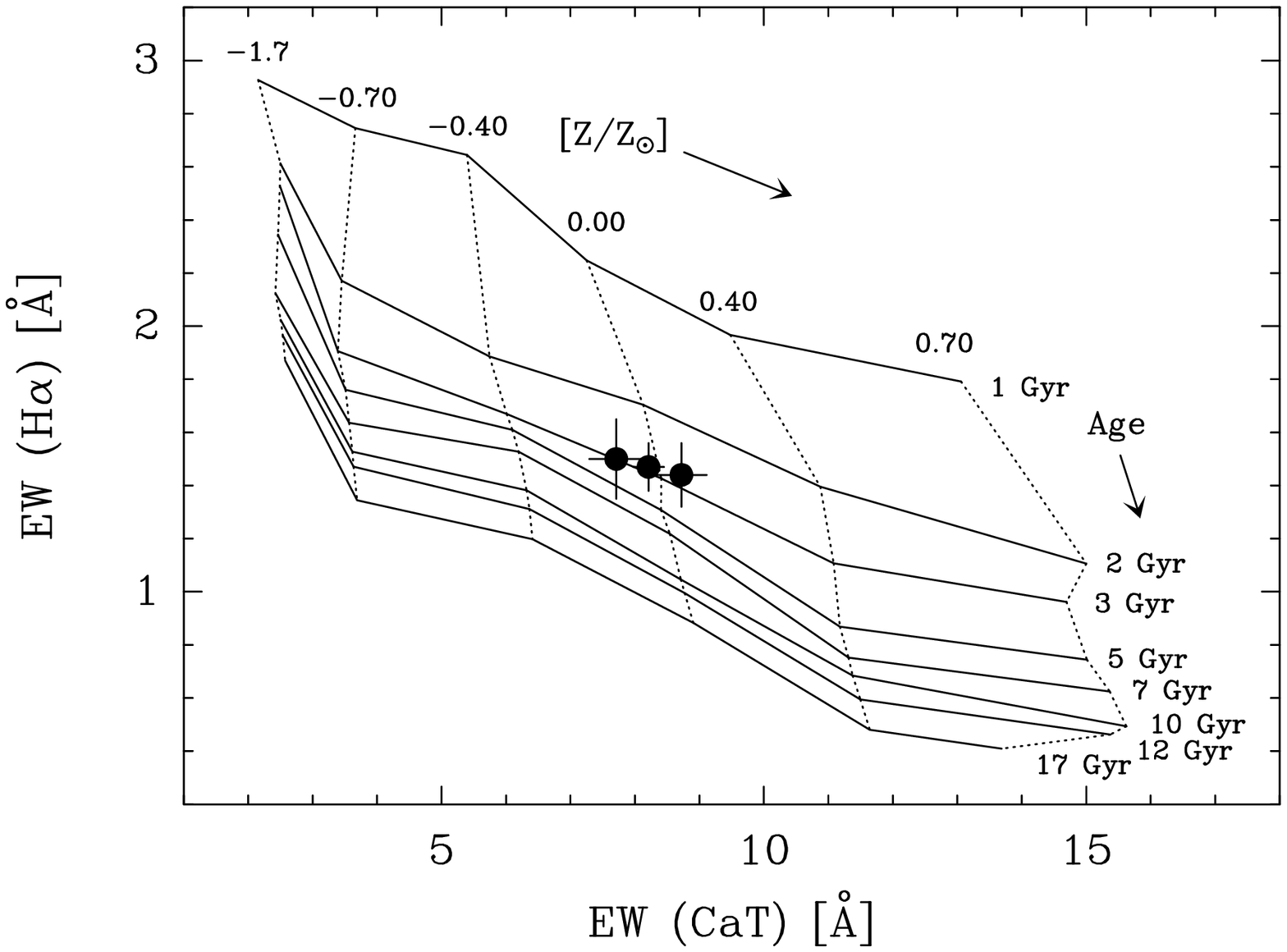,width=7.4cm,angle=270}
}}
\caption{
(a) {\it (left)} H$\beta$\,--\,[MgFe] diagram for two GCs in NGC 7252
(Schweizer \& Seitzer 1998), and
(b) {\it (right)} H$\alpha$\,--\,Ca-triplet diagram for three GCs in
NGC 1316 (Goudfrooij et al.\ 2001a).  The NGC~7252 clusters are $\sim$550~Myr
old, the NGC~1316 clusters $\sim$3.0~Gyr.  Note that both sets of clusters
have near-solar metallicities.
}
\label{fig2}
\end{figure}

The GC system of NGC 7252 has been studied in especially great detail.
Spectroscopy of eight young GCs shows that seven of them feature strong
Balmer absorption lines [EW(H$\beta$)\,= 6\,--\,13~\AA] indicative of a
main-sequence turnoff dominated by A-type stars (Schweizer \& Seitzer 1998).
Six of these GCs have ages in the narrow range 400\,--\,600~Myr.
Figure~\ref{fig1} displays the spectrum of the brightest cluster, W3,
with the Balmer and various metal lines marked.  A first interpretation of
this spectrum with Bruzual \& Charlot (1996) models for an assumed Salpeter
IMF yielded a surprisingly high cluster mass of $1.8\times 10^8\,M_{\sun}$.
However, a reinterpretation of this spectrum with cluster models
incorporating a new treatment of AGB stars yields a lower mass of
$7\times 10^7\,M_{\sun}$, or still $\sim$20 times the mass of $\omega\;$Cen
(Maraston et al.\ 2001). These authors' $K$-band photometry indicates that
most young globulars in the halo of NGC 7252 are presently in their AGB
phase-transition stage (which lasts from 200~Myr to $\sim$1~Gyr).

The metallicities of these young halo GCs appear to be near solar.
Figure~\ref{fig2}a shows a H$\beta$\,--\,[MgFe] diagram for the two
globulars W3 and W6 ({\it data points}), from which $[Z] = 0.00\pm 0.08$
for W3 and $+0.10\pm 0.17$ for W6.  A fascinating object is S101, a very
young halo cluster located in an H$\;$II region that is falling back into
NGC 7252 from a tidal tail and has a metallicity of $[Z] = -0.12\pm 0.05$.
This cluster, located at 15~kpc projected distance from the center,
suggests that young globulars can form with considerable time delays when
tidally ejected gas comes crashing back into a remnant.  

The line-of-sight velocity dispersion of the eight spectroscopically observed
GCs in NGC~7252 is $140\pm 35$~km~s$^{-1}$, leaving little doubt that
these clusters belong to a halo population.  From {\it HST\,} photometry,
there appear to be a few hundred similar young halo GCs in addition to the
old GCs that must have been part of the halo populations of the two input
spirals (Miller et al.\ 1997).  Figure~\ref{fig3} shows a color--magnitude
diagram of all GCs beyond 2~kpc projected distance from the center and the
corresponding color distribution.  Notice in this diagram the uniformly blue
$(V-I)_0$ colors of the luminous young clusters, producing the pronounced
narrow peak at $(V-I)_0 \approx 0.65$ in the color distribution, and the
parallel sequence of fainter, redder clusters, which shows up as a secondary
peak around $(V-I)_0 \approx 0.95$ in the color distribution.

In short, the merger of two gas-rich spirals in NGC 7252 has led to a
young remnant with a clearly bimodal population of halo globular clusters.
Besides the universal old metal-poor GCs the halo now also features a few
hundred second-generation GCs that are {\it young}\, and {\it metal-rich}.
As best as we can tell, the situation is similar in the young remnants
NGC 3597 and NGC 3921, and perhaps also in NGC 1275.  The many properties
that remnants such as NGC 3597, 3921, and 7252 share with ellipticals
suggest not only that these remnants are present-day protoellipticals
(e.g., Schweizer 1998), but also that many E and S0 galaxies with bimodal
cluster distributions may have formed {\it their}\, second-generation,
metal-rich GCs in a similar fashion.  Interestingly, the ratio of young to
old GCs in NGC 7252 is about 0.7 (Miller et al.\ 1997), close to the mean
ratio of metal-rich to metal-poor GCs observed in normal giant ellipticals.

\section{Globular Clusters in Intermediate-Age Merger Remnants}

If indeed E and S0 galaxies with bimodal cluster distributions formed
through similar mergers as those described above, we should be able to
(1) find E and S0 galaxies with second-generation GCs of intermediate age
and (2) trace the evolution of second-generation GC systems from young
through intermediate to old ages.  Potential tracers of such evolution
are, e.g., the GC color distributions, luminosity functions, and
radial distributions.

Evidence for the existence of intermediate-age GCs in elliptical galaxies
has been claimed for about eight systems, including NGC 1316, 5128,
1700, 3610, and 6702 (see Table~1 for refs.).  The best case so
far is NGC 1316 (= For~A), for which there are spectroscopic data to
bolster the claim made from broad-band photometry.  Figure~\ref{fig2}b
shows the equivalent widths of H$\alpha$ for three bright GCs plotted
versus the equivalent widths of their Ca$\:$II triplets. From the superposed
model grid, one can see immediately that all three GCs are about
$3.0\pm 0.5$~Gyr old and have close to solar abundances (Goudfrooij et
al.\ 2001a).  Their ages agree with the ages inferred from $BVI$ and $JHK$
photometry for a larger sample of GCs in this galaxy (Goudfrooij et
al.\ 2001b; see also Goudfrooij, this volume). Therefore, the red peak
of the bimodal color distribution for NGC 1316 clusters clearly contains
GCs of intermediate age, and this merger remnant provides an evolutionary
link between young remnants like NGC 7252 and old ellipticals with
bimodal cluster distributions.

\begin{figure}[t]
\centerline{\psfig{figure=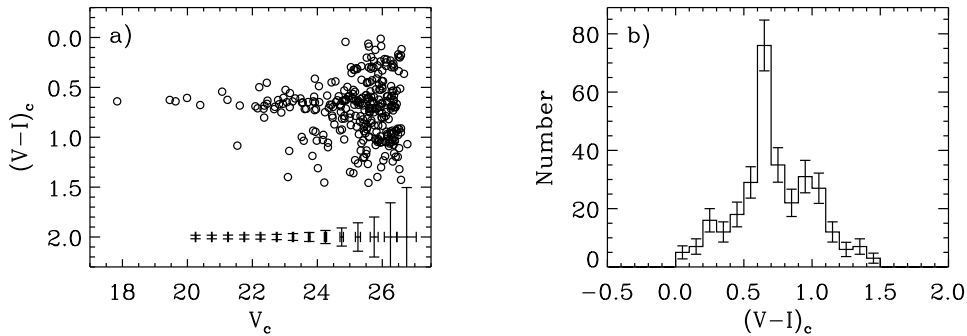,height=4.5cm,angle=90}}
\caption{
(a) $V\!-\!I$ vs.\ $V$ diagram for GCs in NGC 7252 and (b) corresponding
color distribution. (From Miller et al.\ 1997.)
}
\label{fig3}
\end{figure}

\begin{figure}[t]
\centerline{\psfig{figure=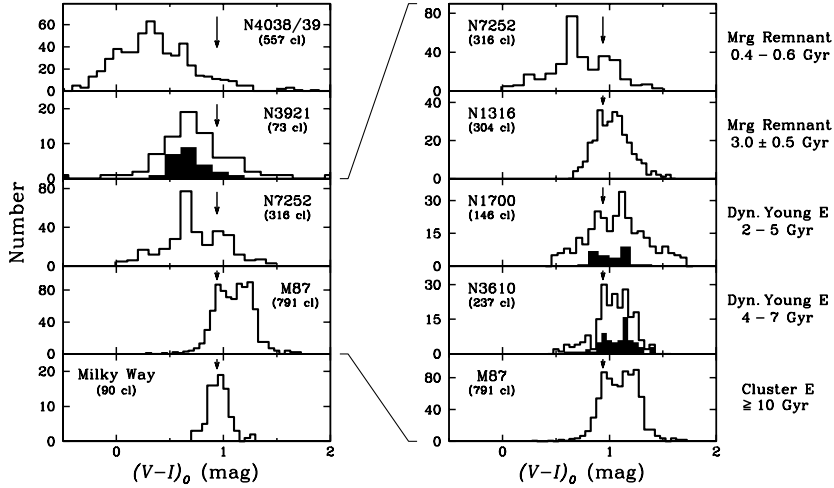,width=11.0cm,angle=-90}}
\caption{
Color distributions of GCs in 3 mergers and 4 ellipticals, compared with
Milky Way.  Location of old metal-poor GCs is indicated by arrows.  Note
color evolution of 2nd-generation GCs from very blue in N\,4038/39 to red
in M87.  The five right panels show that the color crossover occurs between
0.5~Gyr and 3~Gyr (GC ages given in margin).
}
\label{fig4}
\end{figure}

For the other galaxies with candidate intermediate-age GCs we have to rely
on broad-band colors.  Model simulations of bimodal GC
populations with second-generation clusters of solar metallicity suggest
what we might expect to observe at different ages (Whitmore et al.\ 1997,
esp.\ Fig.\ 15):  At 0.5~Gyr the second-generation GCs should appear both
bluer and $\sim$2~mag brighter than the old metal-poor GCs, as observed in
the young remnants discussed above (e.g., Fig.~\ref{fig3}).  At
1.0\,--\,1.5~Gyr they should reach about the same $V-I$ color as old GCs,
but still be $\sim$1.5~mag brighter.  At 3~Gyr they should be already
distinctly redder than the old GCs but still 0.5\,--\,1~mag brighter, while
at $\ga$10~Gyr they will appear both redder and slightly fainter.
Figure~\ref{fig4} illustrates that this predicted crossover of GC colors
indeed occurs.  Shown are the color distributions of clusters in seven
galaxies, with second-generation GCs ranging from very young and blue in
The Antennae to old and red in M87.  Whereas the general evolution from blue
to red colors for second-generation GCs has been known for some time, the
more detailed transition shown in the right-hand panels of Fig.~\ref{fig4} is
new.  The new data for NGC 1316 (Goudfrooij et al.\ 2001b), NGC 1700 (Brown
et al.\ 2000), and NGC 3610 (Whitmore et al.\ 1997 and in prep.) diminish the 
gap in cluster ages from the previous 0.5\,--\,10$^+$~Gyr to currently
about 0.5\,--\,3~Gyr.

\begin{figure}[t]
\centerline{\psfig{figure=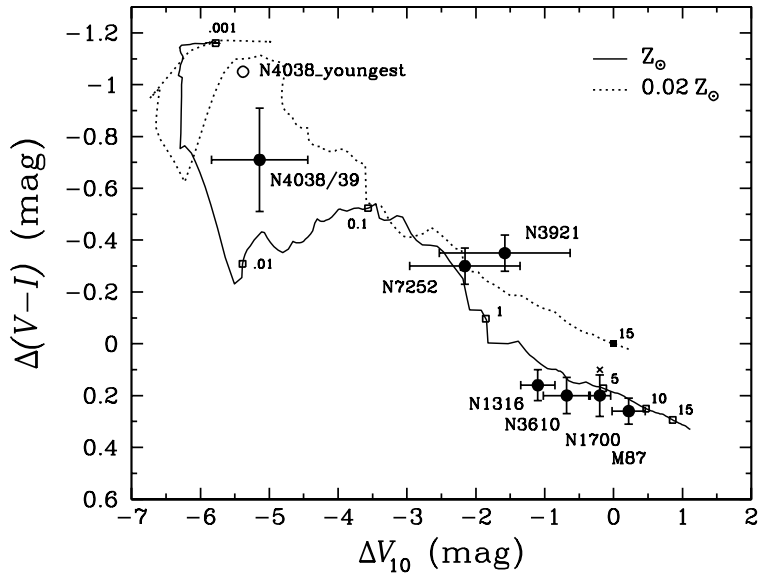,width=10.0cm,angle=270}}
\caption{
$\!\Delta(V\!-\!I)$\,--\,$\Delta V_{10}$ diagram showing color difference
between 2nd-generation and old GCs plotted vs.\ magnitude difference
between 10th-brightest 2nd-generation GC and its old counterpart. Points
with error bars show values for GC systems of 3 merger galaxies and 4 Es.
Lines give evolutionary tracks for model clusters of solar and 1/50th solar
metallicity (Bruzual \& Charlot 1996), and are marked with cluster ages
in Gyr.  Note that the observed seven GC systems form an age sequence.
(After Whitmore et al.\ 1997, with some new data added.)
}
\label{fig5}
\end{figure}

The evolutionary trend as a function of age can be seen even better in the
$\Delta(V\!-\!I)$ vs.\ $\Delta V_{10}$ diagram introduced by Whitmore et
al.\ (1997).  Figure~\ref{fig5} shows such a diagram, in which
$\Delta(V\!-\!I)$, the reddening-corrected color difference between the
peak due to second-generation GCs and that due to old metal-poor GCs, is
plotted versus $\Delta V_{10}\,$, the magnitude difference between the
10th-brightest second-generation GC and the 10th-brightest old GC.  Data
points with error bars give the locations of the GC systems for the seven
galaxies of Fig.~\ref{fig4}.  The data for NGC 1316 are new, as
mentioned above.  For NGC 1700, the old data point by Whitmore et al.\
(marked by a small cross) is superseded by new data from Brown et al.\
(2000), who succeeded in separating the second-generation GCs from the
old GCs.  Note that the GC systems lie roughly along the evolutionary
track for solar-metallicity model clusters ({\it solid line}).  This
supports the notion that most second-generation globulars are relatively
metal-rich ($[Z]\approx -0.8$ to $+$0.2), as verified spectroscopically
for GCs in NGC 7252, NGC 1316, and M87 (Cohen, Blakeslee, \& Ryzhov 1998).
But above all, the $\Delta(V\!-\!I)$\,--\,$\Delta V_{10}$ diagram
demonstrates quite clearly that the GC systems of the three merger galaxies
and four ellipticals form an {\it age sequence}.   

The luminosity function (LF) of second-generation GCs is another potential
tracer of systemic evolution.  The transition from the power-law form
observed in young cluster systems to the log-normal form observed in old
GC systems is now predicted theoretically (Fall, this volume).
It is a consequence of the preferential erosion of low-mass clusters due to
various disruption mechanisms, of which the main one is internal two-body
relaxation and evaporation.  This erosion should be evident in the observed
LFs of second-generation GC systems that form an age sequence.  So far, a
tentative claim has been made only for NGC 1316 (Goudfrooij, this volume),
where the slope of the LF of second-generation GCs appears to be shallower
than the usual $\alpha\approx -1.6$ to $-$2.1\,.

Finally, the radial distribution of GCs within their host galaxies contains
valuable information about both their formation and their dynamical evolution. 
In the young remnants NGC 3921 and NGC 7252 the radial distribution of
second-generation GCs is virtually identical to that of the galaxy light in
$V$ (Schweizer et al.\ 1996; Miller et al.\ 1997).   This indicates that the
young GCs and their progenitors experienced the same violent relaxation as
did the average star, suggesting that the GC progenitors were relatively
compact {\it giant molecular clouds}\, orbiting among the disk stars of the
two input spirals. There is tentative evidence for subsequent central erosion
of GC systems, presumably due to tidal shocking of GCs during passages close
to the center:  At a radius of 1.2~kpc in NGC 1316, the radial distribution of
the GCs shows a deficit of $\sim$45\% relative to the integrated star light
(Goudfrooij et al.\ 2001b), while in old cluster ellipticals the corresponding
deficit is significantly larger still (e.g., Capuzzo-Dolcetta \& Donnarumma
2001). Therefore, as with the color distributions, there appears to be a
continuum of radial distributions of GCs ranging from young-cluster
distributions as strongly centrally concentrated as the host merger remnants
to old-cluster distributions typically less concentrated than the host
ellipticals. It remains to be seen whether this tentative dynamical sequence
will be confirmed as further examples of intermediate-age GC systems are
added to the radial-distribution sample.

In short, second-generation globulars of intermediate age have recently been
found in a few ellipticals and seem to form an evolutionary link between
the young metal-rich GCs observed in recent merger remnants and the old
metal-enriched GCs found in most old ellipticals.  This link is seen most
clearly in the color distributions of GCs, more tentatively in their radial
distributions, and---for the moment---only weakly in their luminosity
functions.

\section{Old Merger Remnants and GC Formation}

Given the propensity of globular clusters to form in the high-pressure
environments of merger-induced starbursts (Schweizer 1987; Jog \& Solomon
1992; Ashman \& Zepf 1992; Elmegreen \& Efremov 1997), GCs in old merger
remnants serve as valuable fossils of the galaxies' early star-formation
history.  The discovery of bimodal cluster populations in ellipticals
(Zepf \& Ashman 1993; Whitmore et al.\ 1995) has convinced many sceptics
that major mergers played a role in forming at least the metal-rich,
second-generation GCs, though alternative formation scenarios have been
proposed as well.  A landmark paper was the observational study of NGC 4472
by Geisler, Lee, \& Kim (1996) who---based on photometry of thousands of
GCs---discovered that the metal-rich clusters lie on average closer to
the center than the metal-poor ones, as predicted by Ashman \& Zepf's
merger model.  This solved the old puzzle of why some GC systems show
steeper radial abundance gradients than their host ellipticals, a fact
attributable in NGC 4472 to the radially varying ratio of metal-rich to
metal-poor GCs.  Mergers involve much gaseous dissipation, which explains
quite naturally the stronger central concentration of second-generation
clusters.

Some recent attempts to age-date metal-poor and metal-rich GCs in cluster
ellipticals have yielded comparable old ages of $\ga$10 Gyr, albeit with error
bars of several Gyr (e.g., Beasley et al.\ 2000; and Kissler-Patig, this
volume).  This begs two questions: (1) How well can we distinguish between
``old'' and ``very old'' ages?  And (2), how much merger activity occurred
during the period of uncertainty that corresponds to the large error bars?

\begin{figure}[t]
\centerline{\psfig{figure=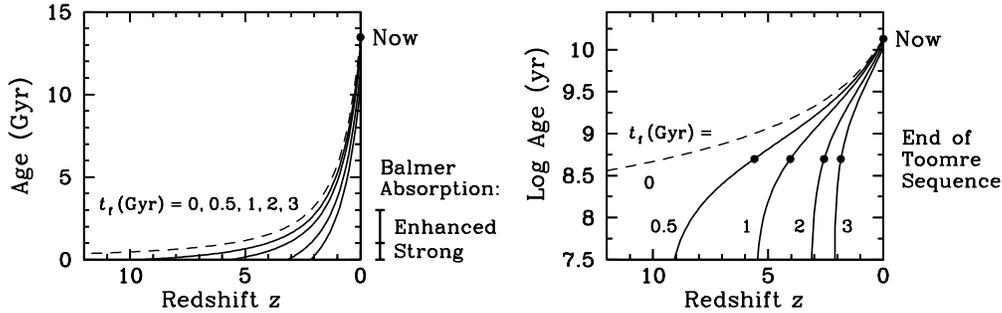,width=13.3cm,angle=270}}
\caption{
Linear ages ({\it left}) and logarithmic ages ({\it right}) of objects formed
at epochs 0, 0.5, 1, 2, and 3~Gyr after Big Bang, plotted vs.\ redshift $z$
in a $\Lambda$CDM cosmology ($H_0$~= 70, $\Omega_{\rm m}$~= 0.3,
$\Omega_{\lambda}$~= 0.7).
}
\label{fig6}
\end{figure}

Figure~\ref{fig6} attempts to address these questions.  Linear and
logarithmic ages of GCs formed at various epochs after the Big Bang are
plotted as a function of redshift $z$ for the currently favored $\Lambda$CDM
cosmology.  Note from the scale bars next to the linear-age plot that the
enhanced Balmer absorption lines that facilitate spectroscopic dating during
the first 3~Gyr fade by $z$~$\approx$~1, even for GCs formed relatively
late at $z_{\,\rm f}$~$\approx$~2.  Alternatively, the logarithmic-age plot
illustrates that all GCs formed during the first 3~Gyr ($z_{\,\rm f}\ga 2$)
appear crunched into $\sim$0.1~dex in age space from our vantage point
(``Now'').  Our current dating techniques for old stellar populations cannot
resolve this tight age interval.  Yet, given the rapidity of the merging
process ($\sim${1/3}~Gyr, see Sec.~1), at least 10 major mergers could have
occurred {\it successively}\, during the universe's first 3~Gyr.  This
illustrates that ``comparable old ages'' for metal-poor and metal-rich GCs
do not in any way argue against the merger model of elliptical and
metal-rich-GC formation.  If metal-poor GCs formed very early on, metal-rich
GCs may well have formed from a few subsequent disk mergers without us
presently being able to measure the 1\,--\,3~Gyr age difference.

Perhaps the single most challenging question concerning GCs is why the
old metal-poor GCs are so universally similar in all types of galaxies
and environments. I believe that observations of GC formation in present-day
mergers and starburst galaxies have yielded a crucial clue:  Whenever an
ensemble of giant molecular clouds is exposed to a rapid pressure increase,
a significant fraction of these clouds get shocked and turn into GCs (Jog \&
Solomon 1992; Fukui, this volume).  The question then is:  Was there a
{\it universal pressure increase}\, early in the history of the universe
that might explain the surprisingly uniform ages and properties of old
metal-poor GCs?

Cen (2001) argues that the cosmological reionization at $z$~$\approx$
7\,--\,15 provided just such a universal pressure increase, which in turn
led to the {\it synchronized formation of metal-poor GCs}\, from early
giant molecular clouds of low metallicity.  If this hypothesis is correct,
the following unified scenario of GC formation emerges.

Most globular clusters in the universe formed from shocked giant molecular
clouds.  The first-generation GCs formed near-simultaneously from pristine
such clouds shocked by the strong pressure increase accompanying
cosmological reionization.  They populate all types of galaxies from
dwarfs through spirals and ellipticals to giant cDs. Later-generation
(``second-generation'') GCs formed during subsequent mergers from
metal-enriched giant molecular clouds present in the merging disks.
{\it Major}\, mergers, some of which occur to the present time, led to
elliptical remnants with a mixture of first- and second-generation GCs
revealed by their bimodal color distributions.  In relatively rare cases
where such mergers occurred near cluster centers, disks stripped of their
gas may have formed ellipticals with mainly first-generation
GCs revealed by unimodal color distributions.  {\it Minor}\, mergers
tended to form S0 galaxies and early-type spirals, again with a mixture of
first- and second-generation GCs.  However, unlike in ellipticals many of
the second-generation metal-rich GCs in disk galaxies may belong to a
thick-disk population if they stem mainly from giant molecular clouds that
belonged to the dominant input disk.  Finally, a minority of
``second-generation'' GCs form sporadically from occasional pressure
increases in calmer environments, such as in interacting Irregulars or
barred galaxies.

\acknowledgements
I thank Bryan Miller for help with Figure 3, and him, Pat Seitzer, Paul
Goudfrooij, and Brad Whitmore for their kind permission to reproduce figures. 
I also gratefully acknowledge research support from the NSF through Grant
AST-99\,00742 and from NASA through various HST grants. 

\begin{table}[ht]
\caption{GCs in Ongoing Mergers and Merger Remnants.\ \ \ }
\label{table1}
\begin{center}\scriptsize
\begin{tabular}{ll}
\tableline
\noalign{\medskip}
Galaxy & References \\
\noalign{\smallskip}
\tableline
\noalign{\medskip}
\noalign{\centerline{ONGOING MERGERS}}
\noalign{\smallskip}
NGC 3256 &              Zepf+ 99                                        \\
NGC 4038/39 (Antennae)& Whitmore \& Schweizer 95; Whitmore+ 99          \\
NGC 6052 &              Holtzman+ 96                                    \\
\noalign{\medskip}
\noalign{\centerline{YOUNG REMNANTS}}
\noalign{\smallskip}
NGC 1275 (Perseus A) &  Holtzman+ 92; Carlson+ 98; {\it Spectra:} Zepf+ 95;
                        Brodie+ 98                                      \\
NGC 3597 &              Lutz 91; Holtzman+ 96; Carlson+ 99; Forbes \& Hau 00 \\
NGC 3921 &              Schweizer 96; Schweizer+ 96                     \\
NGC 7252 (Atoms-for-Peace) &  Schweizer 82; Whitmore+ 93; Miller+ 97; 
			Maraston+ 01; $\phantom{AAA}$ \\
	 &		{\it Spectra:} Schweizer \& Seitzer 93, 98      \\
\noalign{\medskip}
\noalign{\centerline{INTERMEDIATE-AGE REMNANTS}}
\noalign{\smallskip}
NGC 1316 (Fornax A) &   Schweizer 80; Shaya+ 96; Grillmair+ 99;
			Goudfrooij+ 01b; \\
         &		G\'omez+ 01; {\it Spectra:} Goudfrooij+ 01a     \\
NGC 1700 &		Whitmore+ 97; Brown+ 00                         \\
NGC 3610 &              Whitmore+ 97                                    \\
NGC 5018 &              Hilker \& Kissler-Patig 96                      \\
NGC 5128 (Centaurus A)&	Graham \& Phillips 80; Harris+ 84, 92; Zepf \&
			Ashman 93; \\
	 &		Minniti+ 96; Holland+ 99; Rejkuba 01; {\it Spectra:}
			Hesser+ 84, 86                                  \\
NGC 6702 &		Georgakakis+ 01                                 \\
\noalign{\smallskip}
\tableline
\tableline
\end{tabular}
\end{center}
\vskip -0.1cm
\noindent
\baselineskip=6.0pt
{\scriptsize REFERENCES.---Brodie, J.P., et al.\ 1998, \aj, 116, 691;\ \
Brown, R.J.N., et al.\ 2000, \mnras, 317, 406;\ \
Carlson, M.N., et al.\ 1998, \aj, 115, 1778;\ \
ditto 1999, \aj, 117, 1700;\ \
Forbes, D.A., \& Hau, G.K.T.\ 2000, \mnras, 312, 703;\ \
Georgakakis, A.E., et al.\ 2001, \mnras, in press;\ \
G\' omez, M., et al.\ 2001, \aap, in press;\ \
Goudfrooij, P., et al.\ 2001a, \mnras, 322, 645;\ \
ditto 2001b, \mnras, in press;\ \
Graham, J.A., \& Phillips, M.M.\ 1980, \apj, 239, L97;\ \
Grillmair, C.J., et al.\ 1999, \aj, 117, 167;\ \
Harris, G.L.H., et al.\ 1984, \apj, 287, 175;\ \
ditto 1992, \aj, 104, 613;\ \
Hesser, J.E., et al.\ 1984, \apj, 276, 491;\ \
ditto 1986, \apj, 303, L51;\ \
Hilker, M., \& Kissler-Patig, M.\ 1996, \aap, 314, 357;\ \
Holland, S., et al.\ 1999, \aap, 348, 418;\ \
Holtzman, J.A., et al.\ 1992, \aj, 103, 691;\ \
ditto 1996, \aj, 112, 416;\ \
Lutz, D.\ 1991, \aap, 245, 31;\ \
Maraston, C., et al.\ 2001, \aap, 370, 176;\ \
Miller, B.W., et al.\ 1997, \aj, 114, 2381;\ \
Minniti, D., et al.\ 1996, \apj, 467, 221;\ \
Rejkuba, M., 2001, \aap, 369, 812;\ \
Schweizer, F.\ 1980, \apj, 237, 303;\ \
ditto 1982, \apj, 252, 455;\ \
ditto 1996, \aj, 111, 109;\ \
Schweizer, F., et al.\ 1996, \aj, 112, 1839;\ \
Schweizer, F., \& Seitzer, P.\ 1993, \apj, 417, L29;\ \
ditto 1998, \aj, 116, 2206;\ \
Shaya, E.J., et al.\ 1996, \aj, 111, 2212;\ \
Whitmore, B.C., et al.\ 1993, \aj, 106, 1354;\ \
ditto 1997, \aj, 114, 1797;\ \
ditto 1999, \aj, 118, 1551;\ \
Whitmore, B.C., \& Schweizer, F.\ 1995, \aj, 109, 960;\ \
Zepf, S.E., \& Ashman, K.M.\ 1993, \mnras, 264, 611;\ \
Zepf, S.E., et al.\ 1995, \apj, 445, L19;\ \
ditto 1999, \aj, 118, 752.
}
\end{table}



\section*{Discussion}

\noindent {\it Goudfrooij:\,} One minor point for your consideration, on
the metallicity of the GCs in NGC~7252: The gas collapsing to form the
H$\;$II region of approximately solar metallicity in the outer region of
NGC~7252 might have been enriched by, e.g., supernovae occurring when the
massive, 300\,--\,500~Myr old GCs were formed.

\noindent {\it Schweizer:\,} This enrichment scenario seems unlikely.  We
know that the gas clump with the young cluster and H$\;$II region originated
in the outer parts of one of the input spirals, since the farther out
material lies in an interacting disk, the farther out it gets flung into the
tidal tails and the longer it takes to fall back into the remnant.  Therefore,
the gas that formed the 400\,--\,600~Myr old GCs came from more central
regions and should have been somewhat more metal-rich than the currently
returning gas clump, in agreement with the observations.

\noindent {\it Kennicutt:\,} In any picture where the first-generation
clusters form by a universal process at very high redshifts (e.g., the
Cen 2001 paper you cited), wouldn't you expect that the specific frequency
and other properties of old blue clusters would be independent of galaxy
properties?  We have seen evidence earlier in this Symposium that the
metal-poor halo clusters know about the environment in which they formed.
Is this a problem for this picture?

\noindent {\it Schweizer:\,} I do not think that it is a problem.  First,
let me emphasize that in the Cen (2001) scenario only a small fraction of
all giant molecular clouds get transformed into GCs, namely those with very
low angular momentum.  Cen concludes that the clumpy gas is reionized from
outside in:  As the first few generations of stars form, low-density regions
are ionized first and higher-density regions later. Thus, there is more time
for giant molecular clouds in denser regions to self-enrich before they get
compressed and form GCs.  I think this proposed sequence may have caused
the metallicity gap that we are observing in the Milky Way halo, where the
lowest-metallicity stars have [Fe/H]~$\approx$~$-$4 while the
lowest-metallicity globulars have [Fe/H]~$\approx$~$-$2.2.  In more massive
galaxies, the lower abundance limit for GCs may be higher still.

\noindent {\it Forbes:\,} This is a comment.  NGC~1052 shows plenty of
signs of a recent merger, including H$\;$I tidal tails.  Keck imaging
reveals bimodality, but the red GCs are not young.  So although we may
have had a recent gaseous merger, few if any new globulars formed.

\noindent {\it Schweizer:\,} I know of the disturbed H$\;$I disk around
NGC 1052, but not of any claims that NGC 1052 is a {\it recent}\, merger.
From its $U\!BV$ colors, we estimated it to be a 7\,--\,9~Gyr old remnant
(Schweizer \& Seitzer 1992, AJ, 104, 1039).

\end{document}